# Long-distance quantum key distribution in optical fibre

P. A. Hiskett[1], D. Rosenberg[1], C. G. Peterson[1], R. J. Hughes[1], S. Nam[2], A. E. Lita[2], A. J. Miller[3] and J. E. Nordholt[1]

[1]Los Alamos National Laboratory, Los Alamos, New Mexico 87545  [2]National Institute of Standards and Technology, Boulder, Colorado 80305  [3]Albion College, Albion, Michigan 49224

Use of low-noise detectors can both increase the secret bit rate of long-distance quantum key distribution (QKD) and dramatically extend the length of a fibre optic link over which secure key can be distributed. Previous work has demonstrated use of ultra-low-noise transition-edge sensors (TESs) in a QKD system with transmission over 50 km. In this work, we demonstrate the potential of the TESs by successfully generating error-corrected, privacy-amplified key over 148.7 km of dark optical fibre at a mean photon number $\mu = 0.1$, or 184.6 km of dark optical fibre at a mean photon number of 0.5. We have also exchanged secret key over 67.5 km that is secure against powerful photon-number-splitting attacks.

Many classical encryption schemes base their security on the perceived difficulty of efficiently performing certain computational tasks, such as the factoring of large numbers. Quantum key distribution (QKD), on the other hand, allows two users to create a shared, secret, random key for encrypting data, enabling communication that can be proven secure by the laws of physics [1]. Ideally, information is contained in the state of a single quantum, so an eavesdropper ("Eve") is unable to gain information without disturbing the system and revealing her actions. To implement QKD, it is necessary to have a source of single quanta, a method for encoding and decoding information onto and from these quanta, and a protocol for establishing a key. Photons are the obvious choice for sending information over large distances with little decoherence or loss. At present, there are no commercially-available single photon sources, but a heavily attenuated, pulsed laser source provides a practical alternative. Photon statistics from such a laser source follow a Poisson distribution, where the probability of a multi-photon signal is approximately $\mu^2/2$ for mean photon number $\mu < 1$. The presence of these signals must be included in the secrecy analysis of the system, because an eavesdropper could gain information about multi-photon signals without being detected. Hypothetically, in the presence of channel loss an eavesdropper using a sophisticated (but presently unfeasible) photon-number-splitting (PNS) attack [2] could even gain complete knowledge of the key if the mean photon number, $\mu$, exceeds a certain link-loss and therefore distance-dependent maximum value. Such upper limits on $\mu$ set a maximum QKD secret key transmission distance owing to the reduction in signal-to-noise with distance. In this paper we show that maximum secret key transmission distances and rates can be dramatically extended by the use of ultra low-noise transition-edge sensor (TES) single-photon detectors in a novel optical fibre QKD system.

Long-distance fibre-based QKD systems such as the one described in this work usually use phase-encoding. In the "prepare-and-measure" BB84 QKD protocol [3], the sender (Alice) encodes a random bit onto a photon using one of two randomly chosen conjugate bases, and sends it to the receiver (Bob). Bob then performs a measurement on the photon, randomly choosing one of the two possible bases. Their random basis choices are then shared over a public channel and only events where the same bases were used are retained, thereby creating a sifted key. Error correction [4] and privacy amplification [5] are applied to the sifted key to create a shorter, final secret key.

Fibre-based QKD systems usually operate at one of the telecommunications wavelengths where optical fibre has very low loss. Fibre has minimum loss of ~ 0.2 dB/km at 1550 nm, but detector properties play a critical role in the performance of QKD systems and limit the length of a secure link. Most present-day optical fibre QKD systems use InGaAs/InP avalanche photo-diode (APD) detectors operated in Geiger mode, which have excellent timing resolution (<100 ps), but suffer from low efficiencies (~ 20%), usually have large dark-count rates (tens of kHz) and require long dead times (several tens of μs) following photon detection [6], limiting maximum transmission distances to approximately 100 km. In contrast, the TES detectors used in this work [7], can be engineered to have much higher detection efficiency at the target wavelength [8], with much shorter dead-time and have no dark counts, although ambient blackbody radiation creates a background count rate that plays the same ultimate role in a QKD system. Despite present TES timing resolutions of order 100 ns, the high efficiency, low dark count rates, and shorter dead-time of TESs mean that their incorporation in a QKD system can enable key distribution over longer distances, at higher secret bit rates and with higher security. TESs have previously been integrated into a QKD system yielding secret key transmission over 50 km of low-dispersion fibre, and many of the associated experimental details have been discussed [9]. The TESs used in the present work had detection efficiency of 65% at 1550 nm, background count rate of 10 counts per second dominated by blackbody radiation, timing resolution of 90 ns full-width at half-maximum, and dead time of 4 μs.

A simplified schematic of the phase-encoding QKD system is shown in figure 1 and has been discussed in detail in Ref. 10. The system operates at a clock rate of 1 MHz with a single 10 MHz rubidium clock providing synchronization for Alice and Bob.[†] A distributed feedback laser, operating at a wavelength of 1550 nm, is gain-switched to output pulses of width 100 ps. After passing through Alice's phase encoder, which time-multiplexes the signals onto one fibre, the optical signals are attenuated to the single-photon level and coupled into a spool of 202 km of single-mode fibre linked to Bob's phase decoder, which, together with Alice's encoder, comprises a single Mach-Zehnder interferometer. The mean photon number μ of the system is defined as twice the mean photon number of the part of the photon wavepacket that transits the long path of Alice's interferometer, at the point that it leaves her enclave. Alice and Bob encode information and choose measurement bases by applying phases to the photons appropriate for the BB84 protocol [11] using their respective fast electro-optic phase modulators, which are located external to their interferometers [12] for stability. Alice only modulates the phase of the part of the wavepacket that travels through the long path in her phase encoder, while Bob modulates the phase of the part that took her short path. BB84 data is communicated in Bob's detections of photons that take the interfering long-short or short-long paths. Typically in such phase-encoded systems, roughly one half of the transmitted signals yield no information, owing to photons that are either delayed or advanced by several ns relative to the data photons, corresponding to the long-short time difference between the paths in the encoder and decoder. This would create a problem for detectors, such as the TES, that lack sufficient timing resolution to discriminate between the different arrival times. So, our system uses a novel switching technique at the input of Bob's interferometer to eliminate these amplitudes: the photon amplitude for Alice's short (long) path is switched onto Bob's long (short) path, respectively. This allows the TESs to be used in the system and doubles the implementation efficiency of a phase encoded system

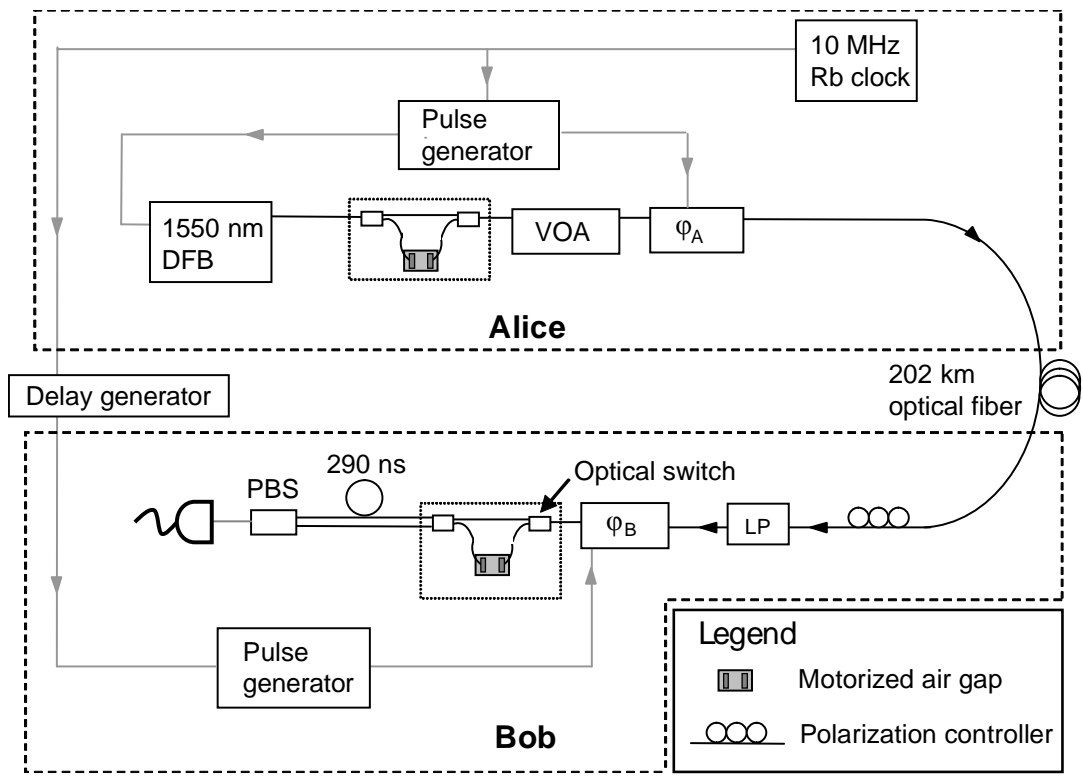

**Figure 1: Simplified schematic of phase-coding QKD system. LP = linear polarizer; VOA = variable optical attenuator; PBS = polarizing beam splitter; DFB = distributed feedback laser.**

By inserting a 290-ns fibre delay into one of the output ports of Bob's phase decoder, and rotating its polarization by 90°, both outputs are time-multiplexed onto a single optical fibre using a polarizing fibre splitter, allowing the receiver to operate with only one TES detector. A histogram of arrival times at the receiver relative to the 1 MHz clock signal displays two peaks, one of which contains events from "0" bits and the other which contains "1" bits after sifting [9]. The peaks are each 90 ns full-width-at-half-maximum and are spaced 290 ns apart. To define the sifted key bits, it is necessary to choose appropriate timing windows for the 0 bits and the 1 bits. Wide timing windows would encompass most of the counts in each channel, maximizing the sifted bit rate, but would also include many background counts, leading to a higher sifted bit error rate (BER). Narrower timing windows would contain fewer background counts, reducing the sifted BER, but would also reduce the sifted bit rate. We chose a width of the timing window to maximize the secret bit rate of the system; this optimal width depends on the rate of real counts compared to the rate of background counts [10]. In general, the optimal window width, which ranged from 30 ns to 170 ns, was narrower for longer distances or lower mean photon numbers.

---

[†] Single-clock synchronization is infeasible in a practical setting outside a laboratory, and a system is under development that uses independent clocks at Alice and Bob.

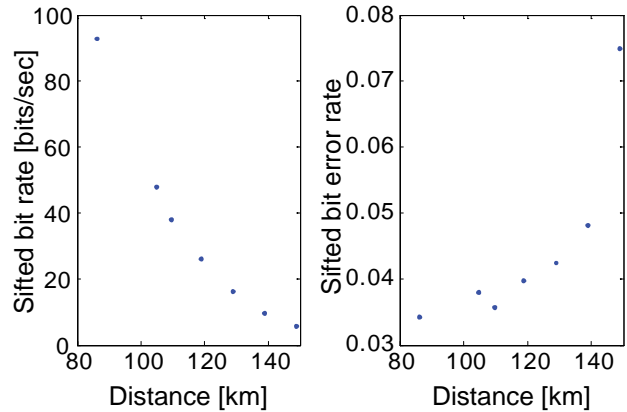

**Figure 2: Sifted bit rate and bit error rate as a function of distance at a mean photon number of $\mu = 0.1$ for optimal window widths. Distances shorter than 202 km were achieved by defining the first part of the fibre link to be within the transmitter's enclave, as discussed in the text.**

Although the length of the fibre link is fixed at 202 km, shorter effective distances can be realized by redefining Alice's transmitter to contain some first portion of the 202 km optical fibre, which acts as an extra attenuator. The mean photon number $\mu$ of signals leaving Alice's enclave must then include the loss in this length of fibre. For example, consider transmissions with $\mu_{202} = 0.5$ at the input to the full 202 km of fibre. Redefining the system so that Alice includes the first 35.8 km of optical fibre, the transmission distance becomes 166.2 km, and we find that $\mu = 0.1$ at the output of the "new" Alice, where we have used the measured attenuation of $\alpha = 0.195$ dB/km for the optical fibre. In general, the relation between effective transmission distance, $d_{eff}$, and mean photon number $\mu$ is

$$d_{eff} = d_0 + (10/\alpha)\log_{10}(\mu/\mu_{202}),\quad(1)$$

where $d_0 = 202$ km. Note that we have only used this technique to map the 202 km fibre link to *shorter* distances; mapping to longer distances would be nontrivial because of effects such as fibre dispersion. In figure 2 we show the sifted bit rate and sifted BER as a function of transmission distance for detection windows optimizing the secret bit rate. The sifted bit rate is consistent with the measured fibre loss, detector efficiency, and window widths, allowing for 7.98 dB loss within Bob's interferometer and optics. The dependence of the BER on window width (not shown) is consistent with the measured background count rate, and from the variation with window width, we infer that the portion of the BER that is solely due to interferometer visibility is 1.8%.

Figure 3 shows the secret bit rate as a function of effective transmission distance of the system, after error correction of the sifted key using the CASCADE algorithm [4], and "BBBSS91" privacy amplification [13] as implemented in Reference [14]. It is assumed that: the Alice-Bob quantum channel losses are random photon deletions with probability corresponding to the measured fibre attenuation; all sifted bits arising from multi-photon signals leaving Alice's enclave are known to Eve; all sifted bit errors are attributed to Eve having performed intercept-resend attacks in the Briedbart basis [13] on single-photon signals in the sifted key; and publicly communicated parity bits for error correction are known to Eve [15]. To facilitate comparison of our results with those of other groups, we have displayed our data at the canonical $\mu = 0.1$ value [13]. We report a new record maximum QKD transmission distance of 148.7 km at this photon number. From a total of 5644 sifted bits we produced 1307 secret bits at a rate of 1.36 secret bit per second (b.p.s) at this distance. However, the choice of $\mu = 0.1$ is arbitrary. Operation at higher $\mu$ yields a higher sifted bit rate and lower sifted BER, but requires more privacy amplification because of the increased likelihood of multi-photon events: for each transmission distance there is an optimal $\mu$ for which the secret bit rate is maximized. In general, it is unlikely for a given system that this optimal $\mu$ is 0.1. If we use equation (1) to map the $\mu = 0.1$ data to a higher $\mu$, the effective transmission distance becomes longer, shifting the $\mu = 0.1$ curve in figure 3 to the right. However, the increased privacy amplification necessary at higher $\mu$ also shifts the curve down, until the secret bit rate for the data point furthest to the right crosses zero. At this point we have reached the maximum transmission distance of our system with BBBSS91 privacy amplification. For our data, the cutoff occurs just over $\mu = 0.5$ and yields a maximum transmission distance of 184.6 km.

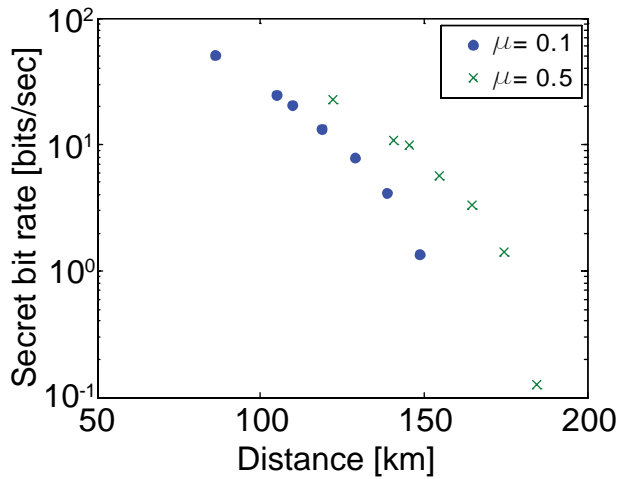

**Figure 3: Secret bit rate as a function of transmission distance, analyzed at $\mu = 0.1$ and at $\mu = 0.5$.**

In the adversarial context of QKD the random-deletion channel assumption of BBBSS91 privacy amplification cannot be rigorously justified with the simple BB84 protocol. For instance, in a PNS attack Eve could hypothetically: block all the single photon signals; remove one photon from each multi-photon signal and store it in a quantum memory; and send the remaining photons from each multi-photon signal over a loss-free channel, to keep Bob's signal detection rate unchanged. Once the bases are announced, Eve could measure her stored photons and gain complete knowledge of the key. Within the simple BB84 protocol, protection against PNS attacks requires operation at mean photon numbers low enough to ensure that at least some of the sifted bits arise from single-photon signals. From 13350 sifted bits with a bit error rate of 5.3%, we have generated secret key secure against general [16] PNS attacks over a 67.5 km fibre link at $\mu = 0.0041$, under the conservative assumption that all of Bob's losses are accessible to Eve.

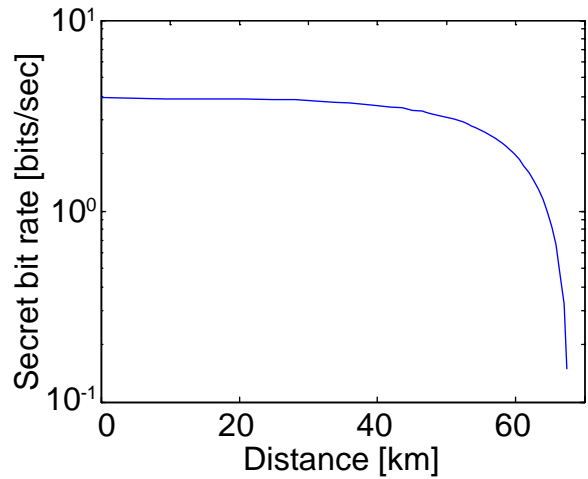

**Figure 4: Secret bit rate as a function of distance for data secure against general photon number splitting attacks.**

Using ultra-low noise, high-efficiency TES detectors in a novel optical fibre QKD system at 1550 nm we have set several new secret key transmission distance records. To the best of our knowledge, at the time this paper was written the distance record for secret key creation in a system with µ = 0.1 was 122 km [17][‡]. We have now increased this record distance by 22% to 148.7 km. The previous record distance for key creation using weak laser pulse QKD with the simple BB84 protocol secure against individual PNS attacks was 50.6 km [19]. We have increased this record distance by more than 30%, and also surpassed by several km the maximum PNS-secure transmission distance inferred in a recent "decoy state" protocol implementation with conventional detectors [20]. Our demonstration of secret key production at 184.6 km at µ = 0.5 under the assumption of a random deletion channel is a new, absolute distance record for QKD. This result indicates that PNS-secure QKD could be extended well into the > 100 km transmission distance regime using TES detectors with a decoy state protocol: the decoy states would provide rigorous justification for the channel properties, without additional assumptions. We observe that our new methodology of using a detection time-window selected to maximize the secret bit rate is likely to be of great value in optimizing the performance of other QKD systems. Finally, we note that significant reductions in TES timing jitter and dead-time are feasible with fairly straightforward improvements in the detector electronics, potentially opening the door to higher secret bit rates over the long transmission distances demonstrated here.

**Acknowledgements**
We would like to thank Alan Migdall for the loan of an optical switch and Joe Dempsey of Corning Inc. for the loan of the 202km of SMF 28e® optical fibre. We note that our measurement of 0.195 dB/km is slightly higher than is expected for SMF 28e® and we attribute this to splices in our system. Jim Harrington is thanked for helpful discussions.  D. R. thanks the DCI postdoctoral program. S.N. acknowledges the support of the DARPA QuIST program and NIST Quantum Inititative.  This work was supported in part by DTO.

---

[‡] Another group has achieved single-photon interference with >80% visibility over a link of 150 km, but their system, which transmitted at µ = 0.2 and did not include phase modulators, was not used to create secret key [18].